# Proposal and Analysis of the Experiment for the Precise Measurement of the Muon Lifetime and the Fermi Constant

A.M. Bakalyarov, A. I. Klimov, I. N. Machulin, E. A. Meleshko, V. I. Selivanov

*NRC Kurchatov Institute, Moscow Russia*

An experiment is proposed for a record-breaking measurement of the world Fermi constant $G_F$ during the decay of a positive muon. Using the GEANT programs, an analysis of all systematic errors was performed. The program "Fast Program" (FP) was developed and tested using GEANT data and running ≈ 1500 times faster. $10^{14}$ events were collected with two options for selecting events. A static accuracy of 0.46 ps for the muon lifetime (currently 2.2 ps) and 0.1 ppm for the $G_F$ value was obtained, taking into account all systematic errors (the experimental accuracy of the $G_F$ value is currently 0.5 ppm). The accuracy of theoretical calculations is 0.14 ppm within the framework of the Standard Model of interactions of elementary particles.

## 1. INTRODUCTION

The world constant $G_F$ describes the decays of elementary particles and nuclei in the weak sector of the Standard Model. In [1], it is the experimental value of $G_F$ that is given as the world constant, since it does not depend on the theoretical model of elementary particle physics. At present, the accuracy of the experimental value $(G_F)/G_F$ = 0.5 ppm [2]. Experiments to measure $G_F$ during muon decay have been carried out 7 times since 1973. by 2013 with an accuracy of $\sigma(G_F)/G_F$ = 68 ppm (1973) and 0.5 ppm (2013) [1]. The error of the theoretical value of $G_F$ in the framework of the Standard Model of interactions of elementary particles is 0.14 ppm [3]. The difference between them is within the error limits. The value of $G_F$ can be measured most accurately during muon decay [3]:

$$\frac{1}{\tau_\mu} = \frac{G_F^2\, m_\mu^5}{192\pi^3}(1+\Delta q),$$

Here $\tau_\mu \approx 2.2$ μs is the muon lifetime, $\Delta q$ describes the correction due to the finite electron mass and the radiative corrections due to the four-fermion interaction in the limit $m_\mu \ll M_W$. Thus, further improvement of the accuracy of the experimental value $G_F$ makes it possible to verify the validity of the Standard Model, for example, the unitarity of the Cabibbo-Kobayashi-Maskawa (CKM) matrix [4], which determines the mixing force in the quark sector and is the source of violation of CP-invariance. CKM, in particular, plays an important role in attempts to explain the dominance of matter over antimatter in the universe. Secondly, this, of course, will improve the accuracy of theoretical calculations. In this paper, we propose and analyze an experimental setup for a record high accuracy measurement of $G_F$. It is shown that it is possible to achieve an accuracy of 0.1 ppm.

## 2. DETECTOR

In the experiment, it is necessary to collect $\approx 10^{14}$ $\mu^+ \to e^+$ decays in a reasonable time of $10^7$ s ≈ 116 days in order to improve the existing accuracy of the muon lifetime $\sigma(\tau_\mu)$ = 2.2 ps ($2\times 10^{12}$ events were collected, [2]). This requires the intensity of the muon beam $\geq 10^7$ $\mu^+$/s. The intensity of the "surface muon" beam at the Paul Scherrer Institute (PSI, Switzerland) reaches $1.5\times 10^8$ $\mu^+$/s [3]. "Surface muons" are formed from pions flying from the surface of a proton accelerator target, and the temporal structure of the beam provides a flow of events that is practically time-independent (Poisson statistics). "Surface muons" have a kinetic energy $E_\mu$ = 4.12 MeV and a track length of ≈ 1.4 mm in a plastic scintillator. The intensity of $10^7$ $\mu^+$/s requires the detector to be divided into many independent parts, since the average time between muons is 0.1 μs, and during 20 μs of observation of $\mu^+ \to$

e⁺ decays, 200 muons and 200 decay positrons appear in one detector. Under these conditions, it is practically impossible to find a positron from the muon that gave birth to it. In addition, the decay background from other muons is at least 200 times greater than the effect. The proposed detector consists of 1024 independent elements (the detector in [2] consisted of 170 elements). This leads to an average time between two muons in one detector of ≈ 100 μs, to a probability of 0.2 of detecting two muons or positrons in the observation interval of 20 μs, and to a low probability of detecting two positrons in our dead time interval of 20 ns (≈ 2 10$^{-4}$). The latter quantity does not depend on the time after the decay of the muon. In [2], this probability was equal to 10$^{-3}$ at the beginning of the registration time and 10$^{-7}$ at the end [5]. When optimizing the detector design, various factors were taken into account: the relative probability of muon diffusion into the neighboring scintillator, the amplitude separation of the muon and positron signals, and the dead time of the silicon photomultiplier (SiPM). The proposed experimental setup is shown in Fig.1.

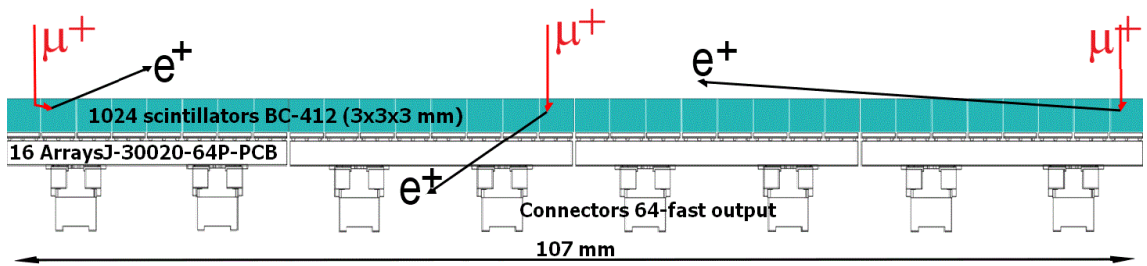

Fig. 1. Sketch of the detector, side view. The scintillators (elements) are separated by 6 μm thick Mylar coated on both sides with aluminum. Beam muons with a momentum $p_μ$ ≈ 28.8 MeV/s and a spread of 2% stop at a depth of 1.4 mm of the BC – 412 scintillators. In most cases, signals from μ⁺ → e⁺ decay particles come from one element. A small percentage of positrons emanate from the neighboring element due to the diffusion of the slow muon, as shown for the muon on the left in the figure. A significant part of the positrons is registered in several elements due to the angular distribution of decay positrons, as shown for the muon on the right in the figure. The complete detector contains 1024 3×3×3 mm³ scintillators, 4×4 = 16 SiPM arrays "ArrayJ-30020-64P-PCB". Each matrix contains 8×8 = 64 SiPM cells with dimensions of 3×3 mm² each. The detector dimensions are 107×107 mm².

The creation of the facility became possible after the development of the technology for mass production of SiPM silicon photomultipliers (for example, [6]), which make it possible to create compact detectors with high speed and a high probability of detecting scintillation photons. On Fig. 2 shows the amplitude distributions of signals with SiPM assuming a leading edge $τ_r$ = 2 ns and a trailing edge $τ_d$ = 4 ns. In this case, the signal is written as ([7])

$$y(t) = R\frac{τ_r+τ_d}{τ_d^2} e^{-\frac{t}{τ_d}}\left(1 - e^{-\frac{t}{τ_r}}\right) \qquad (1),$$

where R is the number of photoelectrons at t → ∞, $y_{max}$ = 0.144R.

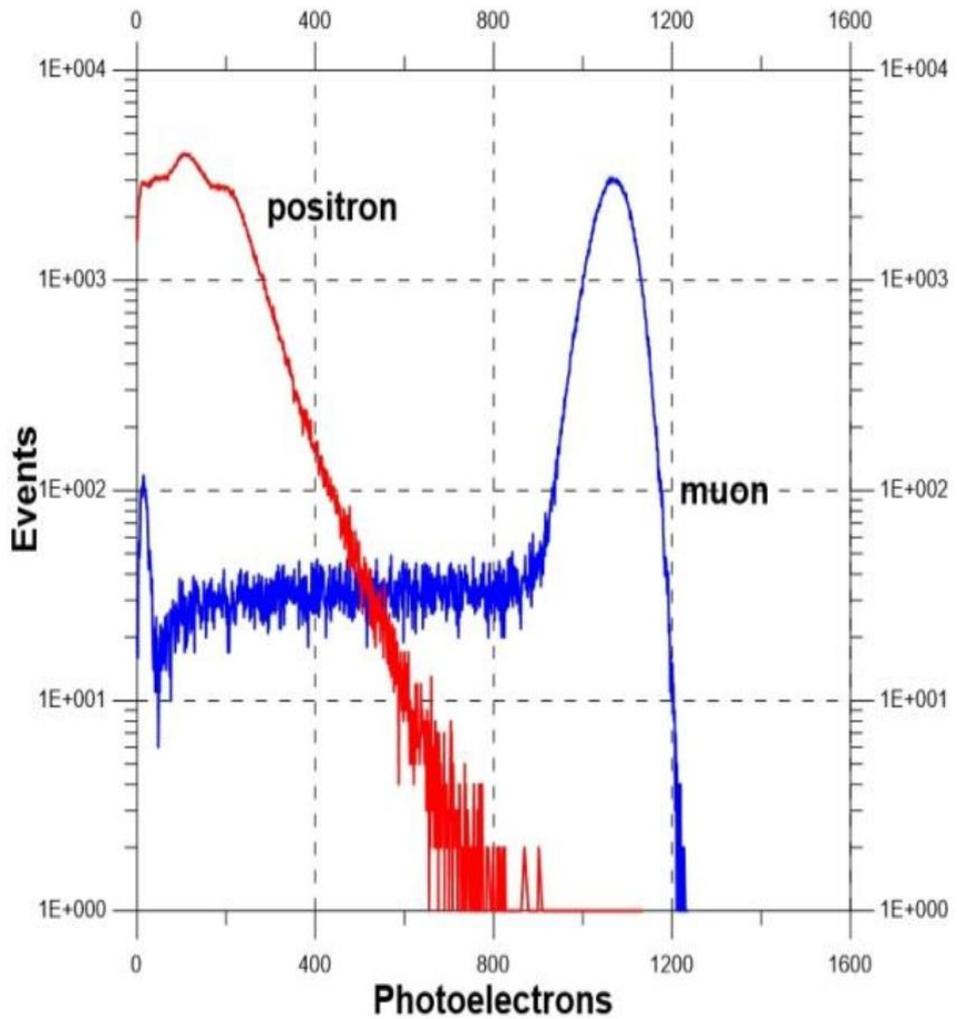

Fig. 2. Spectra of the average amplitude of signals in photoelectrons (phe) from all scintillators at the output of silicon photomultipliers at an intensity of $10^7 s^{-1}$ $\mu^+ \to e^+$ decays in a complete assembly of 32×32 = 1024 crystals. The dimensions of each scintillator are 3×3×3 mm$^3$. The signal spectra for arrays of 3×3 and 32×32 crystals are almost identical. The muon momentum in the beam is 28 MeV/s with a spread of 2%. The spectrum was obtained for the BC - 412 scintillator using GEANT4, taking into account the light yield, its degradation with large ionization losses of the slow muon (Birks effect), light reflections from aluminized Mylar between scintillators, particle stopping in Mylar and the operation of several scintillators due to muon scattering, and due to the angular and energy distributions of positrons. The amplitude spectra are given for the leading edge of the signals $\tau_r$ = 2 ns and the trailing edge $\tau_d$ = 4 ns. Muon signals with an amplitude of < 800 phe arise due to the diffusion of a slow muon into neighboring scintillators. This leads to an error in particle identification at amplitudes < 800 phe. It is possible to improve this identification by measuring the signal amplitudes in all 3×3 = 9 crystals. We abandoned this method because of its errors and significant complication of the experiment. In all further calculations, the particle is identified as a positron in the amplitude range (70 – 800) phe and as a muon in the range (801 – 1200) phe. The minimum amplitude of 70 phe for the positron was chosen with a large margin in order to completely ignore the SiPM noise and its increase during radiation damage.

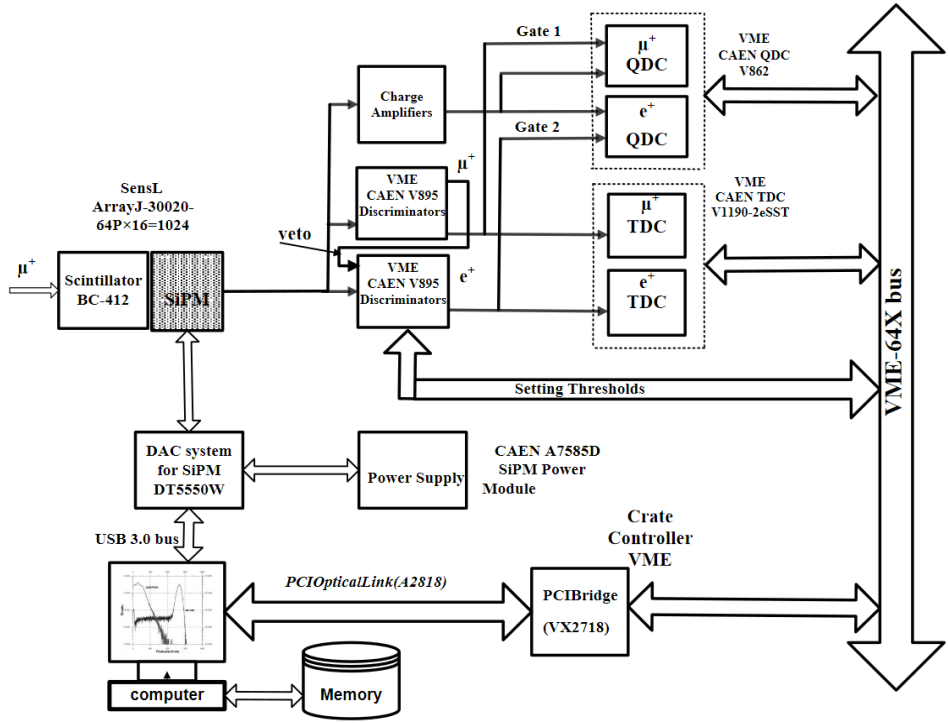

Fig. 3. Possible functional diagram of the equipment for recording signals from SiPM detectors. The devices presented in Fig. 3 are serial [8]. The "veto" signal disables registration of the μ-signal in the e-channel, but this results in random matches between the μ-signal and the second e-signal after the first e-signal. This can lead to spectrum distortions (see 4.1). The probability of such coincidences does not exceed $10^{-7}$ for a dead time of 30 ns for e-signals.

The output signal from one detector element (Fig. 1) is fed to the tracking threshold discriminators (V895) to select the signal from the muon ($\mu^+$) and positron ($e^+$), as well as to the charge-to-code converter (QDC) for the muon signal. Using QDC modules, it is possible to stabilize the gain of the detector for $\mu^+$ and $e^+$ signals during the entire measurement period along the top of the distribution of the number of photoelectrons from muons (Fig. 2). There are several causes of instability - a decrease in the signal due to radiation damage to the scintillator and/or SiPM detectors, a change in amplitude with a change in ambient temperature, etc. Further, the signals are fed to time-to-code converters (TDC). The V1190A/B-2eSST module is a 128/64 channel TDC with a channel width of 0.1–0.8 ns, a measurement range of 104 μs, a dead time of 5 ns for paired pulses, and a 32k × 32-bit output buffer for detecting particles in the observation interval. TDC allows you to efficiently record events with continuous data transfer. The spectrum of the QDC signals is processed and used (sent to the "Power Supply") to change the voltage on the SiPM and thus adjust the gain of the SiPM detectors. Equivalent alternative – adjustment of thresholds of SiPM discriminators ("Setting Thresholds").

### 3. MONTE CARLO

The proposed detector (Fig. 1) was analyzed using the GEANT4 programs (for example, [9]) and the latest version of GEANT3. Electromagnetic processes were analyzed with an accuracy of 1 keV for GEANT4 and 10 keV for GEANT3. Both programs gave identical results. The identity of the GEANT results for 1024 and 9 elements was shown with a statistic of $10^9$ events. This made it possible to collect statistics sufficient for analysis for 9 elements on GEANT in a reasonable time, and the necessary statistics of $10^{14}$ events on the "Fast Program" (FP, see below) for 80 days instead of 516 for a 32×32 assembly.

**3.1. Choice of Beam Intensity, Observation Interval, and Event Selection Criteria**

The dependence of the measurement error of the muon lifetime $\sigma(\tau_\mu)$ on the observation conditions is analyzed. At different intensities, analysis options were considered when choosing one event in the observation interval: 1_μ and/or 1_e or all events: all_μ and/or all_e, as well as options for the presence / absence of anticoincidence between neighboring elements. "1_μ+all_e", for example, corresponds to selecting an event with only one muon and with all positrons in the interval (-30 – +30) μs from the time of the selected muon. "all_μ+all_e" corresponds to the selection of all events in the interval (0 – 20) μs from the time of the selected muon. Table 1 shows the results of this analysis.

Table 1. Results of fitting a GEANT4 assembly of 3×3 crystals with statistics $10^9$ and different muon beam intensities $I_\mu$ for 1024 crystals with different event selection criteria and in the presence of anticoincidence between adjacent elements. The setup logs all events (no dead time). The fitting formula is $N(t)=N_0 \exp(-t/\tau_\mu)+B$. The values of $\chi^2$ lie within natural limits for all variants. It can be seen from the Table that the value of $\sigma_\mu$ increases with increasing beam intensity, apparently due to an increase in background B. In reality, calculations were also performed at intensities $I_\mu = (1.62, 1.89, \text{ and } 2.27) \times 10^7$ μ+/s. All parameters change smoothly from values at $1.26 \times 10^7$ s$^{-1}$ to values at $2.84 \cdot 10^7$ s$^{-1}$.

| Beam intensity $I_\mu$, $10^7$ s$^{-1}$ | Selection at (- 20 – +20) μs | $\Delta\tau_\mu$, (ps) | $\sigma_\mu$, (ps) | $\Delta\tau_\mu/\sigma_\mu$ | $N_0/B$ |
|---|---|---|---|---|---|
| 1.26 | 1_μ+1_e | 444 | 126 | -3.5 | 239 |
|  | 1_μ+all_e | 37 | 133 | 0.3 | 70 |
|  | all_μ+all_e | 41 | 138 | 0.3 | 27 |
| 2.84 | 1_μ+1_e | 972 | 163 | 6.0 | 108 |
|  | 1_μ+all_e | 39 | 166 | 0.24 | 32 |
|  | all_μ+all_e | 118 | 161 | 0.73 | 12 |

The analysis is performed when selecting events without anticoincidence between adjacent elements and with anticoincidence. These cases differ due to positrons being detected in multiple elements and muon diffusion into a neighboring element (see Fig. 1). The results are shown in Table 2.

Table 2. Results of fitting a GEANT4 assembly of 3×3 crystals with a statistic of $10^9$ events, a muon beam intensity of $10^7$ s$^{-1}$ per 1024 crystals with various event selection criteria. "Selection" corresponds to the number of particles in the interval (-20 – +20) μs, but (0 – +20) μs for "all_μ+all_e". The setup logs all events (no dead time). The fitting formula is $N(t) = N_0 \exp(-t/\tau_\mu)+B$. The values of $\chi^2$ lie within natural limits for all variants.

| No anticoincidence between crystals | | | |
|---|---|---|---|
| Selection | 1_μ+1_e | 1_μ+all_e | all_μ+all_e |
| $\sigma(\tau_\mu)$, (мкс) | 0.127 | 0.134 | 0.139 |
| $N_0/B$ | 234 | 70 | 28 |
| 2 ns anticoincidence between crystals | | | |
| $\sigma(\tau_\mu)$, (мкс) | 0.138 | 0.141 | 0.154 |
| $N_0/B$ | 368 | 175 | 36 |

Based on the analysis of the results of Table 2, two options were chosen for modeling the statistics of $10^{14}$ events - the worst ("all_μ+all_e" with anticoincidence) and satisfactory in terms of accuracy ("1_μ+all_e" without anticoincidence). Comparison of the muon lifetime error $\sigma(\tau_\mu)$ shows its decrease by 8% as the registration time increases from 20 μs to 30 μs. Further, this error is unchanged up to the registration time of 120 μs.

On Figures 4 and 5 show the results of fitting the statistics of $10^{14}$ events (3×3 assembly) for the options "1_μ+all_e" and "all_μ+all_e". The beam intensity was $10^7$ μ$^+$/s for a full detector of 1024 elements.

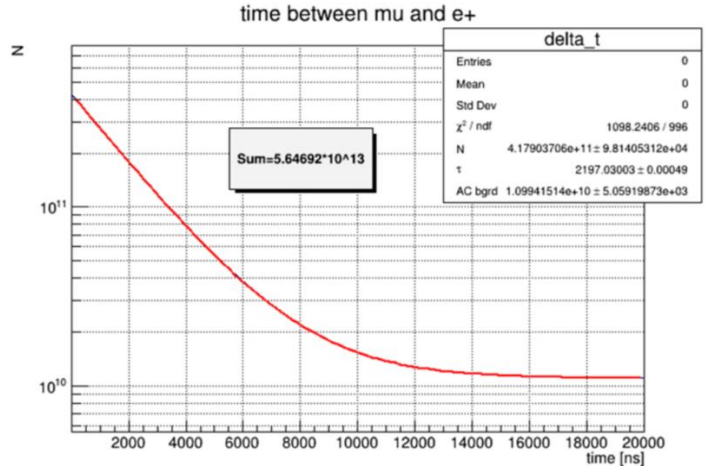

Fig. 4. Time spectrum of positrons during muon decay in the range (0 – 20) μs for $10^{14}$ events. $N(t) = N_0 \exp(-t/\tau_\mu) + AC_{bgrd}$. Assembly of 3×3 scintillators. Program FP (3.2). When simulating the statistics of 1014 events, we obtained $\sigma(\tau_\mu) = 0.48$ ps for the "all_μ+all_e" selection with anticoincidence. When fitting, the value $\tau_0 = 2197.03$ ns was set.

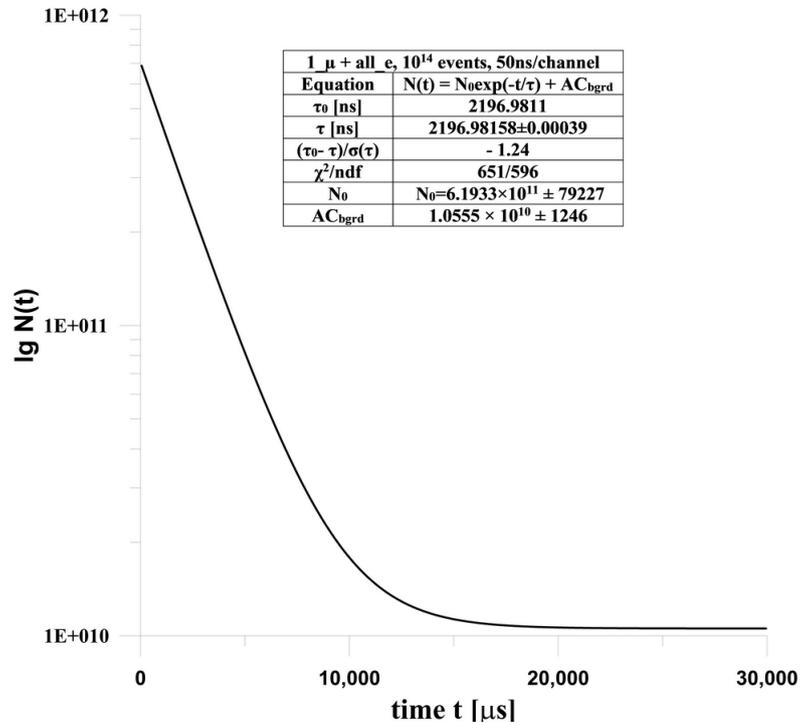

Fig. 5. Time spectrum of positrons during muon decay in the range (0 – +30) μs for $10^{14}$ recorded events. Assembly of 3×3 scintillators. Program FP (3.2). When simulating the statistics of $10^{14}$ events, $\sigma(\tau_\mu)$ = 0.39 ps was obtained for the "1_μ+all_e" variant. Interval (-30 – +30) μs without anticoincidence. When fitting, the value $\tau_0$ = 2196.9811 ns was set.

## 3.2. Fast Program (FP)

To collect $10^{14}$ statistics using GEANT packages would require about 100 years when working on 16 processors. However, to play the whole process in time, it is not necessary to use full modeling of all processes. The time sweep could be modeled using a separate program, which did not have to simulate all physical processes, but it had to contain a whole set (package) of probabilities, which was determined using GEANT complexes on reasonable statistics ($10^9 - 10^{10}$ events) without deployment of the entire process in time. For each muon, the detection probabilities were predetermined in each of the detectors, and possibly in several. The probabilities of registering a positron in one or several detectors and the probabilities in which ones were obtained. We also took into account the probabilities that a muon will be registered in the assembly, but a positron will not be registered, and vice versa, a muon will not be registered, but a positron will be registered. In the second case, the probability of registering a positron in each of the detectors is given. In addition to the above, the probability that a muon will be registered as a positron and vice versa, as well as the probability that there will be no registered particles in the assembly, etc. are also taken into account. The choice of the 3×3 assembly was also due to the fact that this is the minimum assembly that adequately reflects the real picture of events, and it is much easier to form a package of probabilities for it than for the 32×32 assembly. Moreover, additional detectors noticeably slow down the work of a fast program (general increase in iterations in loops, iterations when determining the coincidence time both in different detectors, and when determining events with dead time). This would not allow us to collect the necessary statistics in a given time frame. The set of statistics for 1014 decays takes 80 days for an assembly of 3×3 elements and > 520 days for an assembly of 32×32 elements.

## 4. SYSTEMATIC ERRORS

### 4.1. Dead time of the event registration system (ERS)

In [2], the influence of the dead time of the ERS in the e–e pair on the systematic shift of the experimental value of the muon lifetime $\Delta(\tau_\mu)$ due to the loss of the second positron was studied. Similar systematics is present in the proposed experiment. Figure 3 shows the values of $\Delta(\tau_\mu)$ for the selection all_μ+all_e. In fact, this is only one of the reasons for the shift in $\Delta(\tau_\mu)$.

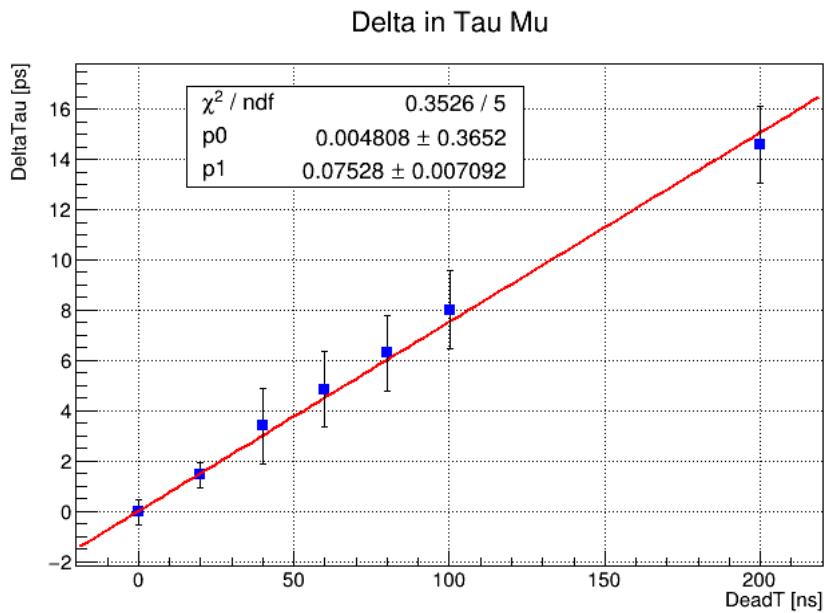

Fig. 6. Dependence of the systematic shift Δ(τ_μ) on the dead time of the ERS. The analysis is given for the "all_μ+all_e" selection option in the interval (0 – 20) μs of muon decay times, while the dead time was equal for registration of μ-e, μ-μ, e-μ and e-e pairs. Anticoincidences are introduced within 2 ns for neighboring crystals. The same analysis was performed with the "1_μ+all_e" selection option. The analysis is given in the interval (0 – 30) μs with respect to the muon detection time only for e–e pairs without anticoincidences between crystals. The shift Δ(τ_μ) in this case is ≈ 7 times larger.

The second reason for the shift Δ(τ_μ) is the 5 ns dead time of the time-to-code converter (TDC) for paired pulses [9]. For example, the minimum dead time in the e-e pair does not exceed 3 ns. The third reason is the shift in the registration time of the second signal in each pair. The fourth one is the dead time of 12 ns of the SiPM amplifying microcell [6]. The dead time and the time shift of the second signal in the pair are calculated for the leading edge of the signal $\tau_r$ = 2 ns and the trailing edge $\tau_d$ = 4 ns (formula (1)) at an amplitude of (70 – 800) phe for a positron and (801 – 1200) phe for a muon. The calculation results are shown in Table 3.

Table 3. {1} is the maximum dead time for registering the second signal and {2} is the maximum time t2 – t1 between the signals in a pair to ensure the shift of the second signal ≤ 1 ps. The leading edge of the signal $\tau_r$ = 2 ns and the trailing edge $\tau_d$ = 4 ns. The positron is (70 – 800) phe, the muon is (801 – 1200) phe. Note that the dead time in the μ-e pair does not affect the systematic shift of the muon lifetime when a time ≥ 100 ns is chosen to detect the decay positron. Threshold tracking discriminators (CAEN V895) with lower threshold were chosen. Discriminators with two fixed thresholds eliminate random matches of the μ-signal with the e-signal (Fig. 3). Discriminators with tracking threshold reduce values in {1} and {2} by 20-30%.

| Particles in pair | μ-μ | μ-e | e-μ | e-e |
|---|---|---|---|---|
| {1}, (ns) | 8 | 25 | 8 | 17 |
| {2}, (ns) | 15 | 45 | 15 | 30 |

Fig. 6 shows the time spectrum N(t2 – t1) of positrons at a dead time of 20 ns only in the e-e pair.

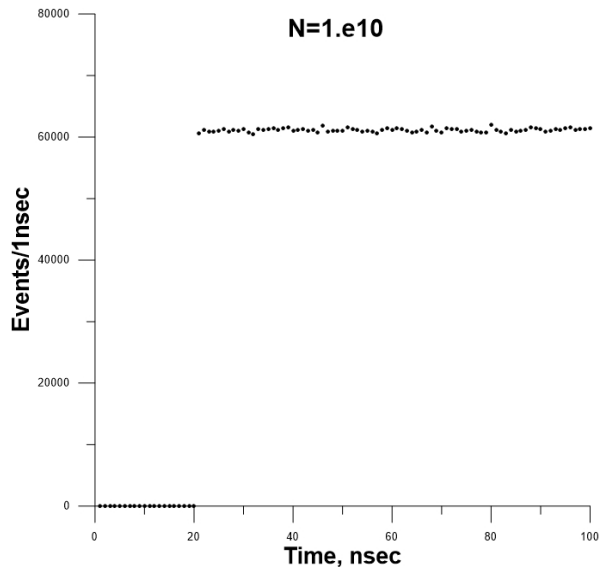

Fig. 6. Dependence N(t_{2e} – t_{1e}) at a given dead time of 20 ns between positrons and at 0 μs < t_{1e} < 29 μs with respect to the muon signal. The value t_{1e} = 0 is assumed in the figure. The variant "1_μ+all_e", there are no anticoincidences between the crystals.

The experiment automatically registers the N(t2 – t1) spectra between all pairs, and you can check the correspondence of the dead times with the results from Table 3. Obviously, the dependences N(t2 – t1) will not be delta-like functions as in Fig. 6 due to the superposition of all four causes and the amplitude spread of the signals, but these dependencies are automatically recorded during the experiment. From Table. 3 it follows that the selection of events with t2 – t1 > 200 ns for pairs μ-μ, e-μ and e-e is guaranteed to exclude the influence of dead time on the value of $\Delta(\tau_\mu)$ and insignificantly reduce the statistics since the probability of such events does not exceed $4 \cdot 10^{-3}$ at an average distance of 100 μs between particles (beam intensity $10^7 \mu^+$/s). The selection of events with t2 – t1 > 100 ns for the pair μ-e will reduce the statistics by only ≈ 5%. The time distributions N(t2 – t1) for all pairs of particles make it possible to measure the dependence of $\Delta(\tau_\mu)$ on the dead time t2 – t1 of any pair and obtain dependences similar to those in Fig. 3. Thus, this makes it possible to exclude the effect of dead time on the accuracy of the value of $\tau_\mu$ with practically no loss in the speed of collecting statistics.

### 4.2. Radiation Damage to Scintillators and SiPMs

Obviously, for a long (116 days) operation time at a beam intensity of $10^7$ μ$^+$/s, it is necessary to analyze the effect of radiation damage on the light output of scintillators, and on the noise and gain of SiPM. In a more difficult position is the scintillator, which is irradiated by muons and positrons. Considerable time was spent searching for a suitable scintillator. The dose per scintillator 3×3×3 mm$^3$ was calculated with the accumulated statistics of $10^{14}$ events per full detector of 32×32 scintillators. It is $2.5 \cdot 10^3$ Gy. In [10], for the ATLAS experiment, the effect of irradiation on the light yield of several scintillators was studied. For the BC-412 (EJ-208) scintillator, a decrease in light output by 10% was observed at a dose of $7 \cdot 10^5$ Gy. It should be noted that the authors used a beam of 6 MeV protons, which have a much greater radiative effect on the scintillator than muons, and at our total dose of $2.5 \cdot 10^3$ Gy there is no change in the BC – 412 light output. It was shown in [11] that a significant decrease in the gain (by 2 times) is observed at a dose of $5 \cdot 10^{10}$ mm$^{-2}$ and is negligibly small at a dose of $5 \cdot 10^8$ mm$^{-2}$. A significant increase in SiPM noise is observed at a dose of $3 \cdot 10^9$ mm$^{-2}$, but this effect is insignificant when choosing a threshold of 70 phe for positron detection.

### 4.3. External magnetic field

We use a muon beam with polarization close to unity [12]. This polarization is partly lost due to the interactions of the muon spin with matter. The polarization of muons in the NE – 102 scintillator was measured in [13]. It is equal to ≈ 0.24. Muon spin will precess in a plane perpendicular to the direction of the magnetic field. This leads to a distortion of the distribution N(t)=N$_0$exp(- t/$\tau_\mu$)+B and the value of $\tau_\mu$ in the presence of a selected geometric direction of positron detection. In the proposed setup, positrons are detected with equal probability in the 4π angle. This needs to be verified as even the Earth's magnetic field ≈ 1 G unacceptably distorts the value of $\tau_\mu$. After that, it is necessary to choose work in a magnetic field >> 1 G or without it. It was shown that the value of $\Delta(\tau_\mu) = 0$, and the values of $\sigma(\tau_\mu)$ are equal at B = 0 and B = 1 kG when fitting the dependence N(t)=N$_0$(1-(P$_\mu$/3)×cosωt)×(exp(- t/$\tau_\mu$)+B) with muon polarization P$_\mu$ = -1 and muon spin precession frequency ω in a magnetic field. Positrons were registered only in the direction of the beam. We will assume that the presence of an external magnetic field is necessary in the presence of a terrestrial field of ≈ 1 G, but this does not affect the results of the experiment.

### 4.4. SiPM Gain Instability

The scatter in the gains of individual cells of the SiPM assembly does not exceed 15% [14]. The change in gain with temperature is ≈ -1%/K. The QDC module (Fig. 3) makes it possible to standardize and stabilize the gain of individual cells during the experiment with great accuracy. We will assume that when these effects are compensated, they will not contribute to the value of $\tau_\mu$.

### 4.5. Time structure of the muon beam

The time structure of the PSI muon beam can be described as $I(t) = I_0(1+ 0.0008 \cdot |\sin(t/40ns)|)$ [15]. It is due to the temporal structure of the proton beam. Statistics were played and processed according to the formula $N(t) = N_0(1 + 0.08 |\sin(t/40ns)|) \times (\exp(-t/\tau_\mu)+B)$ and fitted according to the formula $N(t)=N_0 \exp(-t/\tau_\mu)+B$. The result is $\chi^2/ndf = 19852/19978$ and no shift $\Delta(\tau_\mu)$. Hence it follows that the temporal structure of the PSI accelerator beam does not contribute to the value of $\tau_\mu$.

### 4.6. TDC errors (time-to-code converter)

The main characteristics of TDC are the observation interval T, the channel price LSB (Least Significant Bin), the differential DNL and the integral INL of non-linearity. It was shown that $\sigma(\tau_\mu)$ does not depend on the observation interval at T> 30 μs and on the channel cost at 1 ns <LSB < 100 ns. DNL and INL are also important quantities for collecting statistics of $10^{14}$ events and high accuracy in determining $\tau_\mu$. Consider two ways to measure DNL and INL. The first way is to calibrate the TDC using a calibrated rubidium standard [2]. The authors obtained a systematic error $\sigma_{syst}(\tau_\mu) = 0.13$ ps. This is a small value compared to our statistical error $\sigma(\tau_\mu) = 0.39$ ps. The second way is to feed uncorrelated events to the start and stop inputs of the TDC [16]. In this case, at DNL = INL = 0, the number of events in each channel is the same within the statistical error. DNL ≠ 0 and INL ≠ 0 distort this distribution. The measured values of these deviations allow you to determine the LSB and the correct time of each channel. It was shown in [16] that the DNL and INL dependencies are identical for many TDCs and do not depend on external conditions. With the TDC proposed here, the start is not correlated with the encoder channel because the TDC runs continuously. The values of $\sigma_i(DNL)$ decrease sharply, but the integral nonlinearity remains. For the proposed encoder TDC – V1190A – 2Esst, the nonlinearities are: σ(DNL) <0.2 LSB and σ(INL) <0.3 LSB [8]. It is not clear with what accuracy it will be possible to describe the dependence $t_N = f(N)$ for TDC – V1190A – 2Esst. Therefore, we will pessimistically assume that we will obtain $\sigma_{syst}(\tau_\mu) = 0.13$ ps, as was shown in [2].

### 4.7. Composition of the muon beam

The beam of "surface muons" PSI contains <$10^{-4}$ pions [12] and <1% positrons [17]. Simple estimates show that such a small admixture of pions does not affect the results. Positrons with an energy of ≈30 MeV have a fairly long range and are only scattered in the scintillator and SiPM material. Positrons uncorrelated in time increase the background and can lead to an increase in $\sigma(\tau_\mu)$. There are two ways to study and reduce the effect of positron impurities on the value of $\sigma(\tau_\mu)$. First, collect ≈ $10^{12}$ cases of muon decay per day and extrapolate the obtained value of $\sigma(\tau_\mu)$ to the statistics of $10^{14}$. After that, if the result is unsatisfactory, place a scintillation detector on the beam axis behind the SiPM to detect the positron impurity in the beam. The signal from this detector, coinciding in time with the signal of any SiPM scintillator, will allow such an event to be excluded from the analysis. This leads to random coincidences of the impurity positron with the μ → e decay positron. At an intensity of $10^7$ s$^{-1}$, the average time between decay positrons is ≈ 100 ns. For a real duration of 5 ns of the SiPM output signal and a solid angle of $0.1 \times 4\pi$ of the scintillation detector, this leads to a loss of 5% of the intensity due to random coincidences and an additional 10% of this value due to the solid angle, in total ≈ 0.5%. The variant corresponds to a scintillation detector with dimensions of 10×10 cm$^2$ at a distance of ≈ 35 cm behind the SiPM. The proposed method, as well as the regular statistics collection mode, assumes the synchronous operation of all TDCs. This is possible with the TDC V1190A/B-2Esst, which allows continuous event logging and storage. Such a regime was, in particular, demonstrated in [18].

### 5. CONCLUSION

In this paper, we propose a new setup for accurate measurement of the muon lifetime and the Fermi constant $G_F$ during $\mu^+ \to e^+ \overline{\nu_\mu} \nu_e$ decay and analyze all possible sources of experimental errors. Unlike electromagnetic and strong interactions, all particles participate in weak interactions, the strength of which is determined by the Fermi constant $G_F$. Many processes are determined by the weak interaction. These are the processes of the big bang, nucleosynthesis and combustion of stellar hydrogen, archaeological dating and medical imaging, and fundamental nuclear and particle physics. The value of the constant $G_F$ is best determined by measuring the muon lifetime $\tau_\mu$. First, the purely lepton character of muon decay provides an unambiguous theoretical interpretation. Secondly, accurate measurement of the muon lifetime has become possible at the present time due to the development of methods for accurate time measurement and equipment for collecting the necessary statistics of $\approx 10^{14}$ decays without distortion. It is shown that it is possible to achieve an accuracy of $(G_F)/G_F = 0.1$ ppm. In this proposal of a new experiment, there is no systematic error due to the dead time of the event registration system. In [2], this systematic error exceeded 0.02 ppm. To further improve the measurement accuracy of $G_F$, it will be necessary, in addition to improving the accuracy of $\tau_\mu$, to experimentally improve the accuracy of the muon mass $m_\mu$ (currently $\sigma(m_\mu)/m_\mu = 0.02$ ppm [1] and the contribution to the accuracy of $G_F$ is 0.05 ppm) and improved time calibration (contribution 0.03 ppm, [2]). At present, the accuracy of $G_F$ calculations is 0.14 ppm [3] within the framework of the Standard Model. The main errors are shown in Table 4.

Table 4. Errors of the proposed experiment for the "1_μ+all_e" variant.

| source of error | $\sigma(\tau_\mu)$, ps | $\sigma(G_F)/G_F$, ppm |
|---|---|---|
| muon mass | 0.22 | 0.05 |
| time calibration | 0.13 | 0.03 |
| statistical error | 0.39 | 0.09 |
| total error | 0.46 | 0.10 |
| existing error | 2.2 | 0.5 |

For the "all_μ+all_e" selection option, the total error $\sigma(G_F)/G_F$ does not exceed 0.13 ppm. When processing the results, the following options are available for analysis: all selection options, the introduction of any dead times between any particles, and the choice of a spectrum with or without anticoincidences between any scintillators. Analysis of variants and explanation of their results significantly increases the reliability of the data obtained. After the analysis, the optimal variant can be chosen in terms of the accuracy of $\sigma(G_F)/G_F$ and/or, perhaps, in terms of the value of $\chi^2$.

## 6. ACKNOWLEDGMENTS

The authors are grateful to A.A. Korsheninnikov and M.D. Skorokhvatov for support of the work, and to the Russian Foundation for Basic Research under grant 18-02-00007A which carried out a significant part of the work

______________________________________________________